\documentclass[twocolumn,showpacs,prb,amsmath,amssymb,
superscriptaddress]{revtex4}
\usepackage{graphicx}\usepackage{dcolumn}\usepackage{bm}

\begin{document}

\def\bk{{\bf k}}
\def\bq{{\bf q}}
\def\bQ{{\bf Q}}
\def\ve{\varepsilon}

\title{Optical conductivity of unconventional charge density wave
systems: \\ Role of vertex corrections}
\author{D.N. Aristov}
\altaffiliation[On leave from ]
{Petersburg Nuclear Physics Institute, Gatchina  188300, Russia.}
\affiliation{
Institut f\"ur Theorie der Kondensierten Materie, Universit\"at Karlsruhe, 
76128 Karlsruhe, Germany}
\affiliation{
Max-Planck-Institut f\"ur Festk\"orperforschung, Heisenbergstra\ss e 1,
70569 Stuttgart, Germany}

\author{R. Zeyher}
\affiliation{ 
Max-Planck-Institut f\"ur Festk\"orperforschung, Heisenbergstra\ss e 1,
70569 Stuttgart, Germany}

\date{\today}

\begin{abstract}    
The optical conductivity of a d-CDW conductor is calculated 
for electrons on a square lattice and a nearest-neighbor
charge-charge interaction using the lowest-order conserving approximation.
The spectral properties of the Drude-like peak at low frequencies and
the broad hump due to transitions across the gap at large
frequencies are discussed, also as a function of temperature and of
the second-nearest neighbor hopping term $t'$. We find that vertex
corrections enhance the d.c.\ conductivity, make the Drude peak
narrower and provide a smooth transition from a renormalized 
regime at low to the bare theory at high frequencies. It is also
shown that vertex corrections enhance the temperature dependence
of the restricted optical sum leading to a non-negligible violation of the 
sum rule in the d-CDW state.

\end{abstract}

 \maketitle

%%%%%%%%%%%%%%%%%%%%%%%%%%%%%%%%%%%%%%%%%%%%%%%%%%%%%%%%%%%%%%

\section{Introduction}

The interaction between electrons may give rise to three
different types of order parameters describing superconductivity,
magnetism or a modulation of the charge density. As an example let us
consider the Heisenberg interaction,
\begin{equation}
{\cal H}_H = \frac{J}{2} 
{\sum_{\substack{{<ij>}\\{\alpha\beta\gamma\delta}}}} 
c_{i\alpha}^\dagger {\vec \sigma}_{\alpha\beta} c_{i\beta} 
c_{j\gamma}^\dagger {\vec \sigma}_{\gamma \delta} c_{j \delta}.
\label{HH}
\end{equation}
The sum in Eq.(\ref{HH}) runs over nearest neighbor sites $i$ and $j$,
$\vec \sigma$ is the vector of Pauli matrices, and $c_{i\alpha}^\dagger,
c_{i\alpha}$ are creation and annihilation operators for electrons at
site $i$ with spin projection $\alpha$. In a mean-field treatment
the expectation value of the two creation (or annihilation) operators,
i.e., the order parameter,
may be non-zero which leads to singlet superconductivity with extended s-wave 
or d-wave symmetry or triplet superconductivity with p-wave symmetry.
If the average of two operators from the same site is non-zero we have 
a modulation of finite local spin moments describing a magnetic state.
If the expectation value of two operators with the same spin projection
is non-zero we deal with a charge-density wave (CDW) with an internal symmetry,
because the two operators in the order parameter
belong to different sites. This case will be considered in the following.
For a square lattice with 4 nearest neighbors there are 
four different local order parameters associated with each of the 
four bonds 
which are in general complex valued but also
subject to the condition that the resulting mean-field Hamiltonian must be
Hermitian. The imaginary parts correspond to circulating
currents expressing the breaking of time reversal symmetry. \cite{Hsu}
The order parameters of a CDW may be classified by the point group symmetry 
and have in our case extended s-wave, d-wave or p-wave symmetries. 
Moreover, they show
a modulation throughout the crystal with a wave vector $\bf Q$ corresponding 
to the total momentum of electron-hole pairs. CDW
systems with an internal structure are called unconventional charge-density
waves\cite{Maki1} to distinguish them from conventional ones where the electron
and hole sit at the same site and do not exhibit any internal structure.

Unconventional charge-density wave systems have been invoked to explain
experimental features in organic conductors, \cite{Maki1,Dora1}
2D transition metal dichalcogenides\cite{Neto} and high-T$_c$ superconductors.
\cite{Affleck,Cappelluti1,Benfatto1,Chakravarty1,Iyengar,Oganesyan}
In particular, it has been shown that a charge density wave with d-wave
symmetry (d-CDW or, shortly, DDW) competes with d-wave superconductivity
in the 
$t-J$ model in the large-$N$ limit where $N$ is the number of spin 
projections.\cite{Cappelluti1}
The phase diagram of this model has a quantum critical point (QCP) separating
at $T=0$ the normal phase at large doping from a d-CDW state at lower doping
if superconductivity is disregarded. Allowing also for superconductivity the
QCP separates a pure superconducting state from a ground state where 
superconductivity and a d-CDW coexist. Due to the presence of the d-CDW
the transition temperature for superconductivity exhibits a maximum,
to be identified with optimal doping, and then decreases strongly in the
underdoped regime. The d-CDW state exists above T$_c$ near optimal and
the underdoped region which means that the pseudogap phase should be 
identified with the d-CDW in this model. Various observable quantities
have been calculated within this picture
\cite{Hsu,Cappelluti1,Cappelluti2,Zeyher1,Chakravarty1,Chakravarty2,%
Chakravarty3,Chakravarty4,Greco1,Aristov1,Carbotte1,Aristov2} 
and compared with 
experimental data but no final conclusions have been reached. 

In the following we will study the optical conductivity of a d-CDW system.
In contrast to other recent work\cite{Aristov1,Carbotte1} vertex 
corrections will be taken into account within a conserving
approximation scheme.\cite{Baym} Otherwise a proper definition of the 
current already 
poses a problem. Within a mean-field theory the current can be defined
via the continuity equation for the charge density. The explicit calculation
of the time derivative of the charge density using the mean-field Hamiltonian
of a d-CDW yields the following expression for the longitudinal current in
${\bf k}$ space, \cite{Benfatto2}
\begin{eqnarray}
{\bf j}({\bf q},t) &=& \frac{1}{N} \sum_{{\bf k}\sigma}
\Big[(\nabla_{\bf k} \epsilon_{\bf k}) c_{{\bf k-q}/2\sigma}^\dagger
c_{{\bf k+q}/2\sigma} 
\nonumber \\ &&
-i(\nabla_{\bf k} \Delta_{\bf k}) 
c_{{\bf k-q}/2\sigma}^\dagger c_{{\bf k + Q +q}/2\sigma} \Bigr].
\label{j}
\end{eqnarray}
$\epsilon_{\bf k}$ and $\Delta_{\bf k}$ are one-particle energies and
the d-wave order parameter, respectively. The first term in Eq.(\ref{j})
is the standard expression for the current, the second one
is unconventional and due to the momentum dependence of the order
parameter. The use of the current definition Eq.(\ref{j}) in calculating
the optical conductivity guarantees that the continuity equation is
satisfied\cite{Benfatto2} but it is not apriori clear whether it is at finite
frequencies more correct than the usual procedure based only on the first
term in Eq.(\ref{j}). In particular, one can argue that the exact
interaction term in the Hamiltonian should be used in defining the current.
One would
then obtain only the first term in Eq.(\ref{j}) because the interaction,
for instance, a Heisenberg term, commutes with the density operator. Below
we will show that a correct treatment defines the current only via the
first term in Eq.(\ref{j}) but takes vertex corrections (VC) into account.
In this way the effective current at low-frequency is of the form of
Eq.(\ref{j}) and satisfies the continuity equation whereas 
it resembles more a bare current at high frequencies.

The paper is organized as follows. In the second section we introduce 
some properties of the Hamiltonian of a d-CDW system. In section III
we derive and solve the equation for the current vertex. In section IV
the optical conductivity is calculated and several limiting cases such
as high and low frequencies are considered as well as the weight of the
Drude peak, optical transitions and the optical sum rule. Section V 
contains results from a numerical evaluation of the obtained 
expressions, and we present our conclusions there.

\section{Hamiltonian}
%\widetext
We consider the following Hamiltonian for electrons on a square lattice,
\begin{eqnarray}
{\cal H} &=& \sum_{{\bf k}\sigma} \xi_{\bf k}^{(0)} c_{{\bf k}\sigma}^\dagger
c_{{\bf k}\sigma} 
\nonumber \\  &+&
\frac 12 \sum_ {\substack{{\bf k}{\bf k'}{\bf k''}\\{\sigma\sigma'}
}}
J({\bf k}-{\bf k'}) c_{{\bf k}\sigma}^\dagger c_{{\bf k'}\sigma'}
c_{{\bf k''}\sigma'}^\dagger c_{{\bf k -k'+k''}\sigma}. 
\label{H}
\end{eqnarray}
$c_{{\bf k}\sigma}^\dagger$ and $c_{{\bf k}\sigma}$ are creation and
annihilation operators for electrons with momentum ${\bf k}$ and 
spin projection $\sigma$, respectively. $\xi_{\bf k}^{(0)}$ is equal
to $\epsilon_{\bf k} - \mu$, where $\epsilon_{\bf k}$ are the bare
one-particle energies and $\mu$ the chemical potential. $\epsilon_{\bf k}$
is given by $\epsilon_{\bf k} = -2t({\cos}(k_x) +{\cos}(k_y)) 
+4t'{\cos}(k_x) {\cos}(k_y)$,
where $t$ and $t'$ denote hopping amplitudes between nearest and
next-nearest neighbors, respectively.
The second term
in $\cal H$ is a Heisenberg interaction between nearest neighbors with
$J({\bf k}) = 2J ({\cos}(k_x) + {\cos}(k_y))$, measuring lengths in units of
the lattice constant. From now on we also will measure energy in units of
$2J$ which will simplify subsequent equations considerably.

Extending the spin projections from two to $N$ the ground state of the
above Hamiltonian is a charge-density wave with d-wave symmetry
for some doping regime away from half-filling
and large $N$. The corresponding wave vector is approximately 
${\bf Q} = (\pi,\pi)$. The DDW order parameter is given by
\begin{equation}
\Delta_{\bf k} =\Delta \gamma_d({\bf k}) =
 -2i\gamma_d({\bf k}) \sum_{\sigma'} \int d{\bf k}'\gamma_d({\bf k'})
\langle c_{{\bf k'}+{\bf Q}\sigma'}^\dagger c_{{\bf k'}\sigma'} \rangle,
\label{Delta}
\end{equation}
with $\gamma_d({\bf k}) = ({\cos}(k_x)-{\cos}(k_y))/2$.
Here and below our measure of integration is 
\begin{equation}
d{\bf k} = dk_x dk_y/(2\pi )^2.
\end{equation}
In the following we assume that the parameters are such that
this DDW state is realized and that only the charge channel is relevant in
accordance with the large $N$ limit. It is convenient to introduce
the spinors $\Psi_{{\bf k}\sigma}^\dagger = (c_{{\bf k}\sigma}^\dagger,
c_{{\bf k+Q}\sigma}^\dagger)$. The mean-field part of the Hamiltonian can
then be written as
\begin{equation}
{\cal H}_{MF} = {\sum_{{\bf k},\sigma}}^\prime
\Psi^\dagger_{{\bf k}\sigma} {\widehat H}_{\bk} \Psi_{{\bf k}\sigma}, 
\label{H0}
\end{equation}
with
\begin{equation}
         {\widehat H}_{\bf k} =
       \begin{pmatrix}
        \xi_{\bf k} &  i \Delta_{\bf k} \\
        -i \Delta_{\bf k} &\xi_{{\bf k}+{\bf Q}}
       \end{pmatrix}.
       \label{Ham-mat}
       \end{equation}
The prime at the summation sign means that the sum extends only over 
the magnetic zone which is 1/2 of the original Brillouin zone.
The mean-field contribution of the interaction term in Eq.(\ref{H}) contains 
a diagonal part, which renormalizes $\xi_{\bf k}^{(0)}$ and yields
$\xi_{\bf k}$, and a non-diagonal part, described by $\Delta_{\bf k}$.
For the following it is convenient to introduce the abbreviations
\begin{equation}
\xi_{\pm} = (\xi_{\bf k} \pm \xi_{\bf k+Q})/2,
\label{xi}
\end{equation}
and
\begin{equation}
E_{\bf k} =     \left[ \xi_{-}^2 +\Delta_{\bf k}^2 \right]^{1/2}.
\label{E}
\end{equation}
${\widehat H}_{\bf k}$  
can be diagonalized by the
unitary transformation $U$,
\begin{equation}
U = \exp(i\sigma^1 \theta_{\bf k}/2),
\label{U}
\end{equation}
where $\theta_{\bf k}$ is uniquely defined by the equations
\begin{equation}
\cos{\theta_{\bf k}} =  \xi_{-}/E_{\bf k}, \;\;\;\;
\sin{\theta_{\bf k}} =  \Delta_{\bf k}/E_{\bf k}.
\label{theta}
\end{equation}         
$\sigma^i$  denote Pauli matrices.
We have  $U \widehat H U^\dagger = diag
 (\varepsilon_{1},\varepsilon_{2})\equiv
\widehat h$ and  the new quasiparticle energies are
       \begin{equation}
       \varepsilon_{1,2}= \xi_{+} \pm E_{\bf k}.
\end{equation}
In the following we will deal will non-magnetic states which means that we 
may drop the spin label $\sigma$ and account for sums over spin by 
simply inserting a factor 2.
The fermionic Green's function matrix is given by
       \begin{equation}
       {\widehat G}_{\bf k}(i\omega_n)=
       (i\omega_n -\widehat H )^{-1},
       \label{defmatrG}
       \end{equation}
where $\omega_n$ denotes the (fermionic) Matsubara frequencies $(2n+1)\pi T$. 
$\widehat{G}_{\bf k}$ is diagonalized by the same
matrix $U$, so that ${\widehat G}=U^\dagger \widehat g
U$ with  $ \widehat g = (i\omega_n - \widehat h)^{-1} $.

\section{Vertex equation}

\begin{figure}[!h]
\includegraphics[width=8cm]{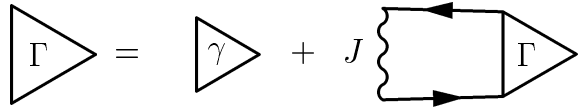}%
\caption{Diagrammatic representation of the integral equation for the 
current vertex
\label{fig:diagrams}}
\end{figure}

The integral equation for the current vertex is depicted graphically in Fig. 
\ref{fig:diagrams} where the two solid lines with arrows
stand for the Green's function matrix
$\widehat{G}_{\bf k}(i\omega_n)$. Analytically, the integral equation reads,
       \begin{eqnarray}
       \widehat\Gamma^{\alpha} (\bq, i \omega_n) &=&
       \widehat\gamma^{\alpha}(\bq) 
         + T \sum_{x_l}\int d\bk\,
       J(\bk- \bq) 
       \nonumber \\ &\times &
       \widehat G_{\bf k}(i x_l)
       \widehat\Gamma^{\alpha} (\bk, i \omega_n)
       \widehat G_{\bf k}(i x_l +  i \omega_n), 
       \label{vertex} 
       \end{eqnarray}
where the current vertex
$\widehat{\Gamma}^{\alpha} (\bq, i \omega_n)$ is a 2x2 matrix with the 
elements $\Gamma^{\alpha}_{ij} (\bq, i \omega_n)$. The bare current
vertex $\widehat{\gamma}^{\alpha}(\bq)$ is diagonal and given by
$\widehat{\gamma}^{\alpha} (\bq) = diag (\partial_\alpha \xi^{(0)}_\bq,
\partial_\alpha \xi^{(0)}_{\bq+{\bf Q}} )$.
The integration over $\bk$ runs over the chemical
Brillouin zone, while $\bq$ lies within the magnetic Brillouin zone. By
choosing this convention we take into account all allowed scattering
processes. Eq.(\ref{vertex}) can also be obtained by rewriting the interaction
term in Nambu notation, using the diagrammatic rules and then extending
the integration over $\bf k$ throughout the chemical and not only the
magnetic Brillouin zone. In Eq.(\ref{vertex}) $i\omega_n= 2\pi i nT$ is the 
(bosonic) external frequency and  the external momentum has been put to zero. We
 neglected   retardation effects on the interaction
amplitude $J(\bk)$ which implies that $\widehat{\Gamma}^\alpha$ depends only on
 one frequency variable.

Performing an analytical continuation Eq.(\ref{vertex}) becomes,

\begin{widetext}
       \begin{eqnarray}
       \widehat\Gamma^{\alpha} (\bq,  \omega) &=&
       \widehat\gamma^{\alpha}(\bq) - \int \frac{dx}{2\pi i}
       d\bk\,  J(\bk- \bq) \left[  n_F (x) \left(
       \widehat G_{\bk}^{A}(x)
             \widehat\Gamma^{\alpha}(\bk, \omega)
        \widehat G_{\bk}^{A}( x +  \omega)
        \right.\right.  \nonumber \\ && \left.\left.
        - \widehat G_{\bk}^{R}(x-\omega)
       \widehat\Gamma^{\alpha}(\bk, \omega)
       \widehat G_{\bk}^{R}( x)  \right)
        + (n_F (x+\omega)-n_F (x))
       \widehat G_{\bk}^{A}(x)
       \widehat\Gamma^{\alpha }(\bk, \omega)
       \widehat G_{\bk}^{R}(x+\omega) \right].
       \label{AnCon}
       \end{eqnarray}
$\widehat{G}_{\bf k}^A$ and $\widehat{G}_{\bf k}^R$ are the advanced and
 retarded 
Green's function matrices, respectively, and $n_F$ the Fermi function.
First we consider the static case, $\omega = 0 $, where the last term 
vanishes.  Inserting $\nabla_{\bf k}^\alpha \widehat{H}_{\bf k}$ for the vertex 
we obtain for the integral on the right-hand side of Eq.(\ref{AnCon}),
       \begin{eqnarray}
       &&\int \frac{dx}{2\pi i}  d\bk\,
       J(\bk- \bq) n_F (x) \left(\widehat G_{\bk}^{A}(x)
       \nabla_\bk ^\alpha \widehat H_{\bk} \widehat G_{\bk}^{A}( x )
       -\widehat G_{\bk}^{R}(x)
       \nabla_\bk ^\alpha \widehat H_{\bk} \widehat G_{\bk}^{R}( x )
       \right)
       \\ &&
       = \int \frac{dx}{2\pi i}  d\bk\,
       J(\bk- \bq) n_F (x) \nabla_\bk^\alpha
       \left( \widehat G_{\bk}^{A}(x)  - \widehat G_{\bk}^{R}(x)\right)
       % \\ &&
       = \nabla_\bq^\alpha   \int dx  d\bk\,
       J(\bk- \bq) n_F (x)\widehat 
       G_{\bk}''(x) =  -\nabla_\bq^\alpha \widehat \Sigma_{\bq},      
       \label{self}
       \end{eqnarray}
\end{widetext}

\noindent where $2 \pi i\widehat G_{\bk}''(x) \equiv\widehat G_{\bk}^{A}(x)  - 
\widehat G_{\bk}^{R}(x) $.        
Noting that the mean-field Hamiltonian is $\nabla_\bq^\alpha \widehat 
H_{\bq} = \widehat\gamma^{\alpha}(\bq) + \nabla_\bq^\alpha \widehat 
\Sigma_{\bq} $,  we observe that (\ref{AnCon})
has the solution\cite{Benfatto2} 
        \begin{equation}
        \widehat\Gamma^{\alpha} (\bq, 0) =
        \nabla_\bq^\alpha \widehat H_{\bq}. 
	\label{static}
        \end{equation}
Eq.(\ref{static}) is a Ward identity relating the 
vertex and the self-energy in the low-frequency limit. 
It is a consequence of the continuity equation for charge and
current densities. Since the self-energy matrix is always non-zero
in the DDW state it means that the current vertex is always renormalized 
and never equal
to the bare one. Only then the continuity equation can be fulfilled at low 
frequencies.
This directly shows the importance of VC in the DDW state. 
However, there may be special cases of interactions where VC
vanish. Such a case has recently been discussed\cite{Carbotte1} 
where $J$ was assumed to have 
the separable
form $J(\bk, \bq) \sim f(\bq) f(\bk)$,  with an even function 
$f(\bq) =f(-\bq)$.  For this  
choice of interaction the integrand in  (\ref{AnCon}) 
is odd in $\bk$, hence the integral vanishes and VC are absent. 
                       
We proceed now to solve the vertex equation Eq.(\ref{vertex}) in the general
case. The second term on the right-hand side of (\ref{vertex}) becomes
after an analytic continuation,
       \begin{eqnarray}
        &&\int d\bk \, J(\bq-\bk) \int dx_1 dx_2 
	\frac{n_F (x_1)- n_F (x_2)}
	{\omega - x_1 +x_2+i0}
        \nonumber \\ && \times
       \widehat G_{\bk}''(x_1)
             \widehat\Gamma^{\alpha}(\bk, \omega)
        \widehat G_{\bk}''(x_2).
       \label{AnCon2}
       \end{eqnarray}
Using the unitary transformation U of Eq.(\ref{U}) the Green's function 
matrix can be
written as $\widehat G_{\bk} = 
U_{\bk}^\dagger \widehat g_{\bk}  U_{\bk}$ with a diagonal $\widehat g_{\bk}$. 
The product of the three matrices in (\ref{AnCon2})
can be represented as 
$U^\dagger \widehat g''(x_1) \widetilde \Gamma^\alpha \widehat g''(x_2) U$ with
$\widetilde \Gamma^\alpha = U \widehat\Gamma^\alpha U^\dagger$. 
Next we expand $\widehat{\Gamma}^\alpha$ in terms of Pauli matrices 
    \begin{equation}
       \widehat \Gamma^\alpha = \sum _i \Gamma^\alpha_i \sigma_i.
       \end{equation}
The $s-$wave component of the current, $\Gamma^\alpha_1$, is in general  
allowed by Eq.(\ref{AnCon}) but vanishes if $J$ is a nearest-neighbor
 interaction.
Hence we restrict ourselves to three components and write 
$\widehat \Gamma^\alpha =  \Gamma^\alpha_0 \sigma_0 + 
\Gamma^\alpha_3 \sigma_3 + \Gamma^\alpha_2 \sigma_2$, and similarly for
the transformed vertex, 
$\widetilde \Gamma^\alpha =  \Gamma^\alpha_0 \sigma_0 + 
\widetilde \Gamma^\alpha_3 \sigma_3 + \widetilde \Gamma^\alpha_2 \sigma_2$. 
Using the explicit form for $U$ we obtain

        \begin{equation}
	\begin{pmatrix}
       \widetilde \Gamma^\alpha_3 \\
       \widetilde \Gamma^\alpha_2
       \end{pmatrix} =	
       \begin{pmatrix}
       \cos \theta_k & - \sin \theta_k \\  
       \sin\theta_k &  \cos  \theta_k 
       \end{pmatrix}              
       \begin{pmatrix}
       \Gamma^\alpha_3 \\
       \Gamma^\alpha_2
       \end{pmatrix}        
       \equiv
       \widehat T        
       \begin{pmatrix}
       \Gamma^\alpha_3 \\
       \Gamma^\alpha_2
       \end{pmatrix}. 
       \end{equation}      
The diagonal components of the expression 
      \begin{equation} 
      I=
     \int dx_1 dx_2 \frac{n_F (x_1)- n_F (x_2)} {\omega - x_1 +x_2+i0}
\widehat g''(x_1) \widetilde \Gamma^\alpha \widehat g''(x_2) 
       \label{kernel}
       \end{equation}
are regularized at $\omega \to 0$ in 
the presence of impurity scattering. We adopt the simplest picture of 
point-like
scatterers, characterized by the inverse quasiparticle lifetime $\tau$. 
Note that isotropic scattering from such impurities allows us to neglect VC
and to include $\tau$ only in the Green's function.
The Green's functions take then the form $\widehat g^{R}(x) = 
diag[g_{1,\bk}^R(x), g_{2,\bk} ^R(x)]$,
  where $g_{j,\bk}^R (x) = 
(x-\ve_{j,\bk}+i/(2\tau))^{-1}$. A standard calculation gives for the 11 
element of $I$
in the limit of small $\tau^{-1} \ll T$, 

       \begin{eqnarray}
       &&(\Gamma_0^\alpha+\tilde{\Gamma}_3^\alpha) \int dx_1 dx_2 
	\frac{n_F (x_1)- n_F (x_2)}
	{\omega - x_1 +x_2+i0}
        g_{1,\bk}''(x_1) g_{1,\bk}''(x_2) 
        \nonumber \\ && \quad = 
	\frac{- n'_F(\ve_{1,\bk})}
        {1-i\omega\tau}(\Gamma_0^\alpha+\tilde{\Gamma}_3^\alpha), 
       \end{eqnarray}
where $n'_F(x)$ is the derivative of the Fermi function.     
The non-diagonal components of $I$
are given by the expression
 \[
 \sigma_2 \frac{\widetilde \Gamma_2^\alpha }{2}
  \left[ 
  \frac{n_F(\ve_{1,\bk}) -n_F(\ve_{2,\bk})}
  {\omega +i/\tau - \ve_{1,\bk} +\ve_{2,\bk} } + (1\leftrightarrow 2)\right].
  \]
Introducing shorthand notations,
       \begin{eqnarray}
        z &\equiv&  (1-i\omega\tau)^{-1} ,\\ 
        \rho_{j} &\equiv& 
        -  n'_F(\ve_{j,\bk}),
        \\
        \rho_{\pm} &=& (\rho_1 \pm \rho_2)/2 , \\ 
       K &=&   \frac{n_F(\ve_{1,\bk}) -n_F(\ve_{2,\bk})}
       {(\ve_{2,\bk} -\ve_{1,\bk})^2-(\omega+i/\tau)^2 } 
       (\ve_{2,\bk} -\ve_{1,\bk}),
       \label{defK}
       \end{eqnarray}
we have 
     \begin{equation}
      I = 
      \sigma_0 z (\rho_+ \Gamma^\alpha_0 +\rho_- \widetilde \Gamma^\alpha_3)
      +\sigma_3 z (\rho_- \Gamma^\alpha_0 +\rho_+ \widetilde \Gamma^\alpha_3) 
      +\sigma_2 K \widetilde \Gamma^\alpha_2. 
      \label{quantityA}
    \end{equation}

The next step in the evaluation of (\ref{AnCon2}) is to carry out the rotation 
$U^\dagger IU$. Suppressing the dependence on the external frequency in the
notation we obtain from Eq.(\ref{vertex}), the following 
system of coupled equations,
\begin{widetext}          
          \begin{eqnarray}
          \Gamma_0^{\alpha}(\bq) &=& 
	  \gamma_0^\alpha(\bq) + z \int d\bk\, 
	  J(\bq-\bk) (\rho_+ \Gamma^\alpha_0 +\rho_- \widetilde \Gamma^\alpha_3)
	  \label{Gamma0}
	  \\    
       \begin{pmatrix} \Gamma^\alpha_3(\bq) \\ \Gamma^\alpha_2 (\bq)
 \end{pmatrix} 
       &=& 
        \begin{pmatrix} \gamma^\alpha_3(\bq) \\ 0 \end{pmatrix}  +
       \int d\bk\, 
	  J(\bq-\bk) 
        \widehat T^\dagger \left[
        \begin{pmatrix} z \rho_- \Gamma^\alpha_0(\bk) \\ 0 \end{pmatrix}        
        +
       \begin{pmatrix} z\rho_+ &0 \\ 0 & K\end{pmatrix} 
       \widehat T 
       \begin{pmatrix} 
       \Gamma^\alpha_3(\bk) \\ \Gamma^\alpha_2 (\bk) 
       \end{pmatrix}
       \right]
       \label{Gamma3}
       \end{eqnarray}

\end{widetext}
 
A further simplification to Eqs.\ (\ref{Gamma0}), (\ref{Gamma3}) 
results from the nearest-neighbor form of the
interaction  $J(\bk)$.
In this case the integral in (\ref{Gamma0}) vanishes, which  
can be seen by considering a shift $\bq \to \bq + 
\bQ$ ;  upon this shift $\gamma^{\alpha}_0$ is unchanged and 
$J(\bk- \bq) \to -J(\bk- \bq)$. 
The ``symmetrical'' component of the current is thus not renormalized 
and given by 
$\Gamma_0^\alpha(\bq) =  \gamma^{\alpha}_0(\bq)= \nabla^\alpha \xi_{+}$.

In order to solve Eq.(\ref{Gamma3}) we note that $\gamma_3^\alpha({\bf q})$ 
transforms like a vector, i.e., according to the representation $\Gamma_5$
of the point group $C_{4v}$. Decomposing $J({\bf k}-{\bf q})$ into separable
kernels only $\sin({k_\alpha})\sin({q_\alpha})$ can contribute because
$\sin(k_\alpha)$ is the only function appearing in the decomposition which
transforms according to the representation $\Gamma_5$. 
Taking
into account the transformation properties of $\widehat{T}$ under $C_{4v}$
one finds that this argument applies also to $\Gamma_2^\alpha$. The solution 
of Eq.(\ref{Gamma3}) has thus the form,
\begin{eqnarray}
\Gamma_3^\alpha({\bf q}) &=& \gamma_3^\alpha({\bf q}) + 
\sin(q_\alpha) c_3^\alpha,
\label{gam1}
\\
\Gamma_2^\alpha({\bf q}) &=&  \sin(q_\alpha) c_2^\alpha.
\label{gam2}
\end{eqnarray}
The coefficients $c_3^\alpha,c_2^\alpha$ depend still on the external
frequency but no longer on momentum.  
It is convenient to introduce the
notation
               
       \begin{eqnarray}
       \widehat{\gamma_0}^\alpha &=& 
       \begin{pmatrix} \rho_{-} \gamma_0^\alpha \\ 
       0 \end{pmatrix},\;\;        
       \widehat{\gamma_3} =  
       \begin{pmatrix} \gamma_3^\alpha   \\ 
       0  \end{pmatrix},
       \label{gam}
\\
       \widehat{V} &=& \widehat{T}^\dagger 
       \begin{pmatrix} z\rho_+ &0 \\ 0 & K\end{pmatrix} 
       \widehat{T}.
       \end{eqnarray}
and 
        \begin{eqnarray}
        \widehat{d} &=& \int d{\bf k} \;\sin(k_\alpha)
        (z\widehat{T}^\dagger \widehat{\gamma}_0 + \widehat{V} \widehat{\gamma}_3 ),
        \label{d}
        \\
        \widehat{X} &=& \int d {\bf k} \;\sin^2(k_\alpha)\widehat{V}.
        \label{XX}
        \end{eqnarray}
Using this notation and the vectors 
        \begin{eqnarray}
       \widehat{c} = 
       \begin{pmatrix} c_3^\alpha \\ 
       c_2^\alpha \end{pmatrix},\;\;        
       \widehat{d} =  
       \begin{pmatrix} d_3^\alpha   \\ 
       d_2^\alpha  \end{pmatrix},
       \label{cc}
       \end{eqnarray}  
one represents Eq. (\ref{Gamma3}) in the following form
        \begin{equation}
        \widehat{c} = \widehat{d} +  \widehat{X} \widehat{c},
        \label{cd}
        \end{equation}

Using Eq.(\ref{theta}) more explicit expressions for the elements of 
$\widehat{V}$ and for $\widehat{d}$ can easily be obtained,
       \begin{subequations}
       \begin{eqnarray}
       V_{11}       
       &=& 
       (z\rho_+ \xi_{-}^2 + K\Delta_{\bf k}^2)/E_{\bf k}^2, \\
       V_{12}       
       &=& 
       V_{21} = 
       \xi_{-}\Delta_{\bf k}(K-z\rho_{+})/E_{\bf k}^2, \\
       V_{22}       
       &=& 
       (z\rho_{+}\Delta_{\bf k}^2 +K \xi_{-}^2)/E_{\bf k}^2,
      \end{eqnarray}
       \label{Xij}      
   \end{subequations}
and
       \begin{subequations}        
         \begin{eqnarray}
       d_3^\alpha
        &=&
        \int d{\bf k} \sin(k_\alpha)\left [
       z\frac{\rho_{-} \gamma_0^\alpha \xi_{-} }{E_{\bf k}}    
        +  {\gamma_3^\alpha}
        V_{11} \right] ,
       \label{d1} 
        \\
       d_2^\alpha
       &=&
        \int d{\bf k} \sin(k_\alpha) \left[
       -z\frac{ \rho_{-} \gamma_0^\alpha
       \Delta_{\bf k}}
       {E_{\bf k}} 
        +  {\gamma_3^\alpha}
        V_{21} \right]. 
        \label{d2}
        \end{eqnarray}                    
       \end{subequations}

\section{Optical conductivity}
              
The frequency-dependent conductivity is given by one current loop
with one renormalized and one bare vertex. Analytically it is
given by the following expression,
     \begin{widetext}
        \begin{equation}
        \sigma(\omega) = \omega^{-1} Im
        \int d\bk\, \int dx_1 dx_2 
        \frac{n_F (x_1)- n_F (x_2)}
        {\omega - x_1 +x_2+i0}
        Tr[\widehat \gamma^{\alpha}(\bk)
        \widehat G_{\bk}''(x_1)
        \widehat\Gamma^{\alpha}(\bk, \omega)
        \widehat G_{\bk}''(x_2)]
        \label{conduc}
        \end{equation}       
Eq.\ (\ref{conduc}) allows the same analysis as was done above for the vertex,
(cf.\ (\ref{AnCon2})). 
The integration over $x_{1,2}$ yields 
        \begin{equation}
        \sigma(\omega) = \omega^{-1} Im
        \int d\bk\, 
        Tr[\widehat \gamma^{\alpha}(\bk)
        U^\dagger I U],
        \label{conduc2}
        \end{equation}    
with $I$ given by (\ref{quantityA}). The integration is over the magnetic 
Brillouin zone. The trace in the last formula is represented as 

   \begin{eqnarray}
   \frac12 Tr[\ldots] &=& z \gamma^{\alpha}_0(\bk)
   (\rho_+ \gamma^\alpha_0 +\rho_- \widetilde \Gamma^\alpha_3)
    +(\gamma^{\alpha}_3(\bk),0)
           \widehat T^\dagger \left[
        \begin{pmatrix} z \rho_- \gamma^\alpha_0(\bk) \\ 0 \end{pmatrix}        
        +
	\begin{pmatrix} z\rho_+ &0 \\ 0 & K\end{pmatrix} 
	\widehat{T} 
       \begin{pmatrix} 
       \Gamma^\alpha_3(\bk) \\ \Gamma^\alpha_2 (\bk) 
       \end{pmatrix}
    \right]. 
    \label{conduc3a}
   \end{eqnarray}  
   \end{widetext}

Eqs.(\ref{conduc2}) and (\ref{conduc3a}) can be recast in the form
\begin{eqnarray}
\sigma(\omega) &=& Im Q/ \omega, \label{conQ} 
\\ 
Q &=& z(a_0 +2a_1) +a_2 + \widehat{d}^\dagger \widehat{c}.
\label{Q}
\end{eqnarray}
The numbers $a_0,a_1,a_2$ are given by
       \begin{subequations}
\begin{eqnarray}
a_0 &=& \int d{\bf k}\;\; (\gamma_0^\alpha)^2 \rho_{+},
\label{a_0} \\ 
a_1 &=& \int d{\bf k}\;\; \widehat{\gamma}_3^\dagger \widehat{T}^\dagger
\widehat{\gamma}_0 = \int d{\bf k} \gamma_3^\alpha \gamma_0^\alpha \rho_{-}
\xi_{-}/E_{\bf k},
\label{a_1} \\ 
a_2 &=& \int d{\bf k} \widehat{\gamma}_3^\dagger \widehat{V} \widehat{\gamma}_3
=\int d {\bf k} (\gamma_3^\alpha)^2 V_{11}.
\label{a_2}
\end{eqnarray}
       \end{subequations}
The dagger at the column vectors denote the transposed vectors, i.e.,
row vectors. Solving Eq.\ (\ref{cd}) we obtain from Eq.\ (\ref{Q})
\begin{equation}
Q = z(a_0+2a_1) + a_2 + \widehat{d}^\dagger (1- \widehat{X})^{-1} \widehat{d}.
\label{QQ}
\end{equation} 
The frequency dependence of $Q$ is due to the functions $F$ and $K$ which
also enter $\widehat{V}$ and thus $\widehat{d}$, $\widehat{X}$ and the vertex.

\subsection{Conductivity at zero and at high frequencies}

In the limit $\omega \to 0$, the conductivity may be determined from the 
relation 
      \begin{equation}
      \sigma(\omega=0) = \tau  (dQ/dz)\rvert_{z=1}.
      \end{equation}
Writing $a_2 = za_{21}+a_{22}$ and $\widehat{d} = z\widehat{D}_1+\widehat{D}_2$
we obtain with $A = a_0 + 2a_1 +a_{21}$,
      \begin{equation}
      \sigma(\omega=0) = \tau  (
      A +\widehat{D}_1^\dagger \widehat{c} +\widehat{d}^\dagger 
      \widehat{c}'  ) \rvert_{z=1},
      \label{eq:tmp1}
      \end{equation}
where ${\widehat c}' =  (d{\widehat c}/dz)$ 
Differentiating Eq.(\ref{cd}) with respect to $z$,
\begin{equation}
(1-\widehat{X}) \widehat{c} = \widehat{d},
\label{ccd}
\end{equation}
and writing $\widehat{X} = z\widehat{X}_1 + \widehat{X}_2$ we obtain
\begin{equation}
(1-\widehat{X}) \widehat{c}' = \widehat{D}_1 +\widehat{X}_1\widehat{c}. 
\label{cs}
\end{equation}
Multiplying Eq.(\ref{ccd}) from the left with $\widehat{c}'^\dagger$ and using
Eq.(\ref{cs}) we find for the last term in Eq.(\ref{eq:tmp1}),
\begin{equation}
\widehat{d}^\dagger \widehat{c}' = \widehat{D}_1^\dagger \widehat{c} + \widehat{c}^\dagger
\widehat{X}_1^\dagger \widehat{c}.
\label{dc}
\end{equation}
Using the abbreviation $\widehat{c}_0 \equiv \widehat{c}\rvert_{z=1}$ we have
\begin{equation}
\sigma (\omega = 0) =
\tau (A +2 \widehat{D}_1^\dagger \widehat{c}_0
+\widehat{c}_0^\dagger \widehat{X}_1 \widehat{c}_0)
 .
\label{sig}
\end{equation}
The last expression is not very transparent, but one can check that it 
exactly corresponds to the naive symmetrization of (\ref{conduc}) using
the static mean-field vertex $\widehat{\Gamma}^\alpha({\bf k},0)$,    
        \begin{eqnarray}
        \sigma(\omega)_{naive} &=&  \omega^{-1} Im
        \int d\bk\, \int dx_1 dx_2 
        \frac{n_F (x_1)- n_F (x_2)}
        {\omega - x_1 +x_2+i0} 
        \nonumber \\ && \times
        Tr[\widehat \Gamma^{\alpha}(\bk, 0)
        \widehat G_{\bk}''(x_1)
        \widehat\Gamma^{\alpha}(\bk, 0)
        \widehat G_{\bk}''(x_2)],
        \label{conducNaive}
        \end{eqnarray}    
in the limit $\omega\to 0$. This result for a simpler case without 
DDW order is usually attributed to Langer. \cite{Langer62}

Consider now the ``tail'' of the Drude peak
which corresponds to $\omega\tau \gg 1$ or $|z| \ll 1$ and $K \equiv K(0)$.
From Eq.(\ref{Q}) we obtain in linear order in $z$,
        \begin{eqnarray}
        Q &=& z(A+\widehat{D}_1^\dagger (1-\widehat{X}_2)^{-1}\widehat{D}_2
        \nonumber \\
        &+ &\widehat{D}_2^\dagger(1-\widehat{X}_2)^{-1}\widehat{X}_1
        (1-\widehat{X}_2)^{-1}\widehat{D}_2 
        \\
        &+&\widehat{D}_2^\dagger(1-\widehat{X}_2)^{-1}\widehat{D}_1).
        \nonumber 
        \end{eqnarray}
Using $\widehat{c} \rvert_{z=0} = (1-\widehat{X}_2)^{-1} \widehat{D}_2$
we find,
        \begin{equation}
        \sigma(\omega \tau \gg 1) \simeq \frac 1{\omega^2\tau}(A
        +2\widehat{D}_1^\dagger \widehat{c} + \widehat{c}^\dagger \widehat{X}_1 
        \widehat{c}) \rvert_{z=0}.
        \label{sigma}
        \end{equation}
Eqs.(\ref{d1}) and (\ref{d2}) show that the two last terms in Eq.(\ref{sigma})
are roughly by a factor $J\Delta_0^2/t^3$ smaller than the first three terms.
Since in practice $J\Delta_0^2/t^3 \ll 1$ we may drop the last two terms
in Eq.(\ref{sigma}) and obtain approximately,
        \begin{equation}
        \sigma(\omega \tau \gg 1) \simeq \frac {A}{\omega^2\tau}.
         \label{sigma1}
        \end{equation}

Eq.(\ref{sigma1}) shows that the current vertex changes from its 
mean-field value Eq.(\ref{static}) near the maximum of the Drude peak to its 
bare value in the tail of the Drude peak. The renormalization 
of the current vertex is important at low frequencies where it contributes
to satisfy the continuity equation of charge. At higher frequencies this
renormalization is stripped off and the bare value of the vertex determines
the behavior in the Drude tail. This observation is not surprizing, 
because the overall prefactor $\tau^{-1}$ shows that impurity scattering 
contributes in the lowest order. The inclusion of the scattering 
process into the corresponding current loop effectively breaks the 
diagram into 
two unconnected parts with their vertices. It is easy to understand that these 
vertices are described by i) small but finite incoming frequency $\omega$, ii) 
the full inclusion of the $J$ term, and iii) 
the Green functions without damping
$\tau^{-1}$. The result of the vertex summation in this limit is given by 
solving (\ref{cd}) at $z=0$ which leads to $\widehat \Gamma^\alpha
\simeq \widehat \gamma^\alpha$ in leading order in $\Delta_0/t$.

\subsection{Weight of the Drude peak}

The weight of the Drude peak is given by integration of the above $\sigma$ 
over $\omega$ for $K(\omega) \equiv K(0)$. 
We define the spectral weight of the Drude peak as $W_D = 
\pi^{-1} \int d\omega Im Q/\omega$. The function $Q(z)$ is an analytical 
function of the complex variable $z=1/(1-i\omega \tau)$ in the circle 
$|z-1/2| < 1/2$, which 
corresponds 
to analyticity in the upper semiplane of $\omega$. Performing a 
Taylor expansion around $z=0$ 
(a more accurate expansion around $z=1/2$ leads to the same result), we write 
$Q(z) = \sum_{n\geq 0} Q_n z^n$ and obtain,

      \begin{eqnarray}
      W_D &=&  \frac 1\pi \int \frac{d\omega}{\omega} Im \sum_{n\geq 0} 
      \frac{Q_n}{(1-i\omega\tau)^n} 
      \nonumber \\ 
      &=& \sum_{n\geq 1} {Q_n} = Q(z=1) -Q(z=0).
      \end{eqnarray} 
From Eq.(\ref{Q}) we have
\begin{eqnarray}
W_D &= &  A + R,
\label{WD}  \\ 
R &=& (\widehat{D}_1^\dagger + \widehat{D}_2^\dagger)(1-\widehat{X}_1 
-\widehat{X}_2)^{-1} (\widehat{D}_1 +\widehat{D}_2) 
\nonumber \\
&& - \widehat{D}_2^\dagger 
(1-\widehat{X}_2)^{-1} \widehat{D}_2. \label{R}
\end{eqnarray}
Inserting the identity
\begin{eqnarray}
(1-\widehat{X}_1 -\widehat{X}_2)^{-1} &=&  
(1-\widehat{X}_2)^{-1}
\nonumber \\ 
&& +(1-\widehat{X}_2)^{-1}\widehat{X}_1
(1-\widehat{X}_1 -\widehat{X}_2)^{-1},
\nonumber 
\end{eqnarray}
into the first term of $R$ and using
\begin{eqnarray}
&&(1-\widehat{X}_1 -\widehat{X}_2)^{-1} \widehat{d} = \widehat{c}_0,
\label{Hi1}
\\&& 
(1-\widehat{X}_2)^{-1} \widehat{d} = \widehat{c}_0 -(1-\widehat{X}_2)^{-1}
\widehat{X}_1 \widehat{c}_0,
\label{Hi2}
\end{eqnarray} 
we obtain
\begin{equation}
R = \widehat{D}_1^\dagger \widehat{c}_0 + \widehat{D}_2^\dagger (1-\widehat{X}_2)^{-1}
(\widehat{D}_1 + \widehat{X}_1\widehat{c}_0).
\label{R2}
\end{equation}
Writing $\widehat{D}_2^\dagger = \widehat{d}^\dagger -\widehat{D}_1^\dagger$
in the second term on the right-hand side of Eq.(\ref{R2}), using the transposed
equation of Eq.(\ref{Hi2}), and inserting the result into Eq.(\ref{WD})
one finds
\begin{eqnarray}
W_D &=&  A +2\widehat{D}_1^\dagger \widehat{c}_0 +\widehat{c}_0^\dagger 
\widehat{X}_1 \widehat{c}_0 
\nonumber \\ &-&
(\widehat{c}_0^\dagger \widehat{X}_1 +\widehat{D}_1^\dagger)
(1-\widehat{X}_2)^{-1} (\widehat{D}_1 + \widehat{X}_1 \widehat{c}_0).
\label{WD2}
\end{eqnarray}
Without the last term in Eq.(\ref{WD2}) $W_D$ would be equal to the 
spectral weight of a simple Lorentzian of the form 
$ \sigma_D(\omega) = \sigma(0)/(1+\omega^2\tau^2) $ 
with $\sigma(0)$ given by Eq.(\ref{sig}). From Eq.(\ref{Xij}) follows that
$1-X_{ij}$ is a positive definite matrix for $\Delta_0/t \ll 1$. Stability
arguments suggest that this should be generally true as long as the DDW state
is thermodynamically stable. It follows that the last term in Eq.(\ref{WD2})
is negative and thus reduces the spectral weight compared to that of a 
Lorentzian form for $\sigma(\omega)$. 

\subsection{Optical transitions}

At finite frequencies of the order of $\Delta_0$ the Drude contribution
to $\sigma(\omega)$ is small and can be neglected whereas the frequency
dependence of $K(\omega)$ becomes important. Putting $z=0$ and showing the
frequency dependencies explicitly Eqs.(\ref{cd}) and (\ref{Q}) take the form,
        \begin{eqnarray}
       \widehat c &=& {\widehat D}_2(\omega) + \widehat{X}_2(\omega) \widehat{c} 
       \label{main1finite}
       \\
       Q &=& 
       a_{22}(\omega) + \widehat{D}_2^\dagger (\omega) \widehat{c}.
       \label{main2finite}
       \end{eqnarray}
If ${\widehat X}_2$,$\widehat{d}_2$ and $a_{22}$ are purely real,
then  $\sigma(\omega) = Im Q/\omega =0$. This situation happens at 
$T=0$, when $\omega$ lies within the optical gap of $K(\omega)$. At
frequencies above the threshold value, which is twice the value of the gap at 
the hot spots, i.e., $\omega \agt 2\Delta_{hs}$, the optical  
conductivity is finite. From Eqs.(\ref{main1finite}) and (\ref{main2finite})
follows that 
\begin{equation}
Q = a_{22}(\omega) + \widehat{D}_2^\dagger (\omega) (1-\widehat{X}_2(\omega))^{-1}
\widehat{D}_2(\omega).
\label{Qdyn}
\end{equation}
The second term on the right-hand side of Eq.(\ref{Qdyn}) is due to 
VC and has been neglected in a previous work.\cite{Aristov1}

\subsection{Restricted optical sum} 
       
Finally we discuss the restricted optical sum, given by the
expression $W = \pi^{-1}\int _{-\infty}^\infty d\omega Im Q/\omega$. 
From the Kramers-Kronig relations we find
\begin{equation}
W = Q(\omega=0) - Q(\omega = \infty).
\end{equation}
The substraction of the term $Q(\omega = \infty)$ reflects the fact that the 
Kramers-Kronig relation recovers the real part of the function up to a 
constant, which is $Q(\omega = \infty)$. The latter quantity is given by 
taking the limit $\omega\to \infty$ in Eq.(\ref{Qdyn}). From the explicit
expressions Eqs.(\ref{Xij})-(\ref{d2}) one finds $Q(\omega = \infty) = 0$.
Furthermore we have,
\begin{equation}
Q(\omega = 0) \equiv Q(z=1) =(A +a_{22} 
+\widehat{d}^\dagger \widehat{c})\rvert _{z=1}.
\end{equation}
Performing a similar calculation as in the case of the Drude weight
we find
\begin{equation}
W = W_D + (a_{22} + \widehat{D}_2^\dagger (1-\widehat{X}_2)^{-1} \widehat{D}_2)
\rvert_{z=1}.
\label{W}
\end{equation}
Comparing with Eq.(\ref{Qdyn}) one recognizes that the sum of the second and 
third terms on the right-hand side of Eq.(\ref{W}) represents just the 
spectral weight due to optical transitions as it should be.

\section{Numerical results and conclusions}

Introducing the unperturbed Hamiltonian $\widehat{H}_{\bf k}^{(0)}$ by
\begin{equation}
         {\widehat H}_{\bf k}^{(0)} =
       \begin{pmatrix}
        \xi_{\bf k}^{(0)} &     0         \\
               0           &\xi_{{\bf k}+{\bf Q}}^{(0)}
       \end{pmatrix},
       \label{H00}
       \end{equation}
the self-energy $\widehat{\Sigma}_{\bf k}$ can be written as
\begin{equation}
         \widehat{\Sigma}_{\bf k} = \widehat{H}_{\bf k} -
         \widehat{H}_{\bf k}^{(0)} =
       \begin{pmatrix}
        \delta \xi_{\bf k}    &i\Delta_{\bf k}            \\
       -i\Delta_{\bf k}       &\delta \xi_{{\bf k}+{\bf Q}}
       \end{pmatrix}.
       \label{Sigma}
       \end{equation}
After dropping the derivative $\nabla_{\bf k}^\alpha$,  
Eq.(\ref{self}) represents a self-consistent matrix equation for
$\widehat{\Sigma}_{\bf k}$. The resulting gap equation reads,
using $2J$ as the energy unit, 
\begin{equation}
1=-2\int d{\bf k} \; \gamma_d^2({\bf k}) \frac{n_F(\epsilon_{1{\bf k}})
-n_F(\epsilon_{2{\bf k}})}{\epsilon_{1{\bf k}}-\epsilon_{2{\bf k}}}.
\label{ggap}
\end{equation}
The renormalization of the diagonal part of 
$\widehat{\Sigma}_{\bf k}$ becomes $\delta \xi_{\bf k}^+ = 0$ and,
writing $\delta \xi_{\bf k}^-=b\gamma_s({\bf k})$,
$\gamma_s =\frac 1 2 ({\cos}(k_x) +{\cos}(k_y))$,
\begin{equation}
b = -2\int d{\bf k} \;\gamma_s({\bf k}) \xi_{\bf k}^-
\frac{n_F(\epsilon_{1{\bf k}}) -n_F(\epsilon_{2{\bf k}})}
{\epsilon_{1{\bf k}}-\epsilon_{2{\bf k}}}.
\label{b}
\end{equation}

We have solved Eqs.(\ref{ggap}) and(\ref{b}) by iteration. For 
$T=0,t=0.25,t'=0.075$ we find that $\Delta$ is non-zero in the interval
$0.35 < n_{\sigma} < 0.50$, increases linearly in this interval from
$0.15$ to $0.19$, and shows a step-like behavior at the ends of this interval.
$n_{\sigma}$ is the occupation of one site for one spin direction.
$b$ is practically constant in the above interval and equal to $-0.39$. 
\begin{figure}[htb]
\includegraphics[width=8cm]{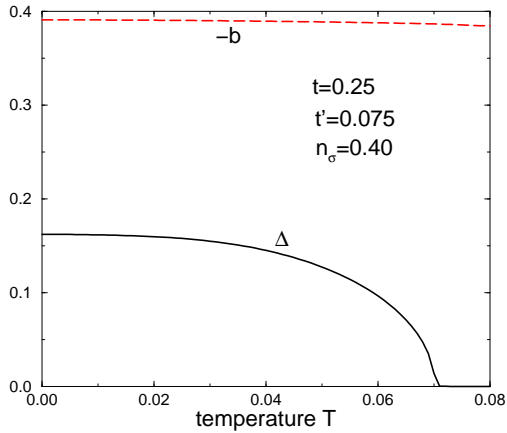}%
\caption{Spectral gap $\Delta_0$ and the parameter $b$ as a function of 
temperature $T$, using $2J$ as the energy unit.
\label{fig:par}}
\end{figure}
Fig.\ref{fig:par} shows $\Delta$ and $b$ as a function of temperature for a 
fixed occupancy $n_\sigma = 0.40$. $b$ is again practically
a constant whereas $\Delta$ exhibits a BCS-like temperature dependence
with $T_c \sim 0.07$. 

\begin{figure}
\includegraphics[width=8cm]{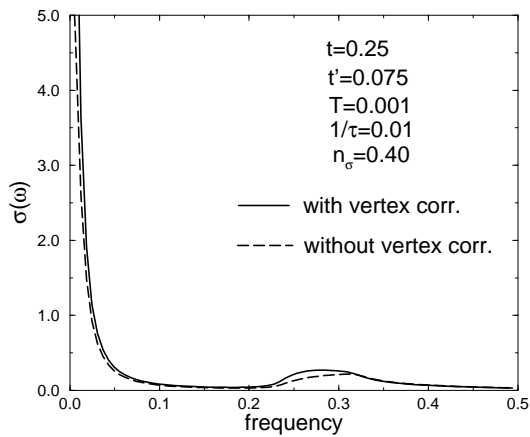}%
\caption{Optical conductivity $\sigma(\omega)$ with (solid line) and
without (dashed line) vertex corrections.
\label{fig:cond}}
\end{figure}

Using the above values for $\Delta$ and $b$ Fig.\ref{fig:cond} 
shows the optical
conductivity $\sigma(\omega)$ calculated with (solid line) and
without (dashed line) VC for the parameters indicated in
the figure. ``Without VC '' means the use of the bare
vertex, i.e., to put $c_2^\alpha$ and $c_3^\alpha$ in Eqs.(\ref{gam1})
and (\ref{gam2}) to zero. 
In the presence of a finite $\Delta$ the Fermi lines
are folded back into the small magnetic Brillouin zone, a gap opens
near the new phase boundary in ${\bf k}$ space, and the resulting
Fermi lines consist of arcs around the nodal direction and small pieces
near the points $X=(\pi,0)$ and $Y=(0,\pi)$. The low-frequency Drude-like
behavior of $\sigma(\omega)$ is due to electrons near the remnant Fermi lines,
especially near the arcs around the diagonal directions.\cite{Aristov2} The 
hump at frequencies above $0.2$ is caused by vertical transitions between the 
two back-folded bands in the magnetic Brillouin zone. These optical transitions 
set in near $2\Delta$ and form a broad and extended continuum due to 
transitions also far away from the Fermi lines.

\begin{figure}[htb]
\includegraphics[width=8cm]{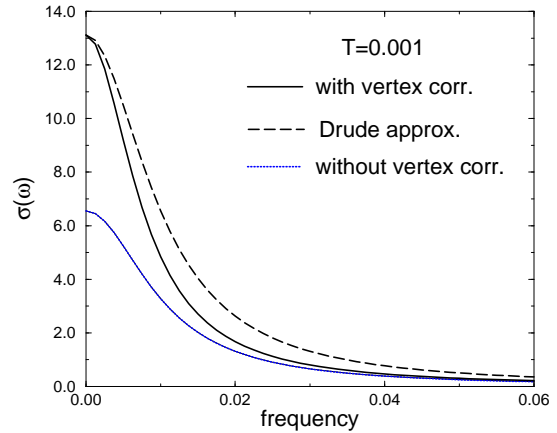}%
\caption{Low-frequency behavior of $\sigma(\omega)$ with (solid curve) and
without (dotted curve) vertex corrections. The dashed curve is a 
Lorentzian with the same value $\sigma(0)$ as the solid curve.   
\label{fig:Drude}}
\end{figure}

In the optical region VC increase 
$\sigma(\omega)$ slightly at frequencies not too far away from $2\Delta$.
This increase agrees with Eq.(\ref{Qdyn}) because the second term on the 
right-hand side is due to VC and it is positive because
the elements of $\widehat{X}_2$ are small compared to unity. The influence of
VC at small frequencies is illustrated in greater detail
in Fig.\ref{fig:Drude}. 
First, VC increase the static value $\sigma(0)$ 
substantially. One reason for this is that the continuity equation for
the charge density yields in the presence of a momentum-dependent
order parameter an additional contribution to the current, given by
the second term in Eq.(\ref{j}). However, $\sigma(0)$ is not only below
but also above $T_c$ enhanced by VC. This increase is due to the 
renormalization
of the one-particle energies. 
The resulting increase of $\sigma(0)$
has in our approach to be ascribed to VC because only  
a non-trivial
vertex allows to fulfill the continuity equation, see Eq.(\ref{static}).
Secondly, the dashed curve in Fig.\ref{fig:Drude}, 
representing a Lorentzian with the same 
$\sigma(0)$ as for the solid line, illustrates the narrowing of the
Drude peak with increasing frequency. This effect is induced by the 
frequency dependence 
of the vertex. The figure shows that the solid line approaches at larger
frequencies the curve without VC (dotted line) which is a 
Lorentzian. This crossover takes place at the impurity scattering
rate $1/\tau$.  
Fig.\ref{fig:Drude} also shows that the effect of VC
cannot be simulated in a theory without VC by a suitable
redefinition of the current. The use of the bare current (first term in 
Eq.(\ref{j})) would result in a violation of the continuity equation and miss
the enhancement of $\sigma(\omega)$ at low frequencies. On the other hand,
the use of a
current derived from the continuity equation as in Eq.(\ref{j}) would
yield the dashed curve in Fig.\ref{fig:Drude} and would thus 
overestimate $\sigma(\omega)$
at large frequencies. 
\begin{figure}[htb]
\includegraphics[width=8cm]{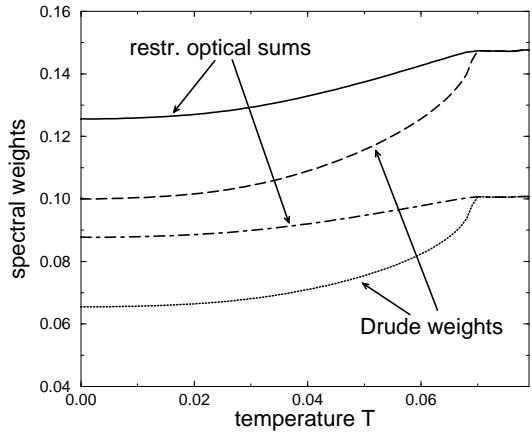}%
\caption{Restricted optical sums with (solid line) and without 
(dot-dashed line)
vertex corrections, and Drude weights with (dashed line) and without
(dotted line) vertex corrections.
\label{fig:weights}}
\end{figure}

Fig.\ref{fig:weights} shows the temperature dependence of the restricted 
optical sum
$W$ and its low-frequency part, the Drude weight $W_D$. VC
affect $W$ in a two-fold way: They enhance the weak temperature
dependence of $W$, calculated without VC (dot-dashed line),
and also increase substantially the absolute value of $W$. The optical
spectral weight $W-W_{D}$ is expected to set in below $T_c$ in a square-root
manner and to be rather independent of VC. Thus the
enhancement of $W$ and its stronger temperature dependence by VC
should be related to the low-frequency part of $\sigma(\omega)$,
i.e., to $W_D$, which is in agreement with the dashed and dotted curves 
in Fig.\ref{fig:weights}. 

\begin{figure}[htb]
\includegraphics[width=8cm]{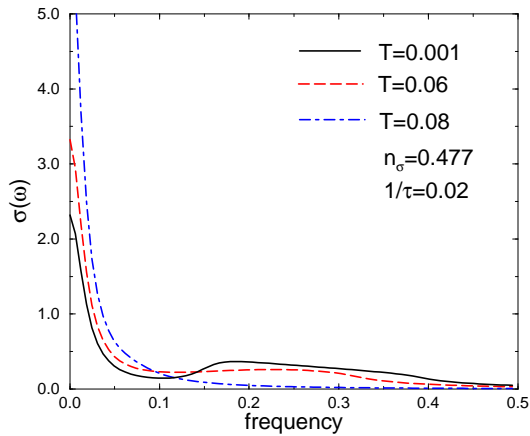}%
\caption{Optical conductivity $\sigma(\omega)$ for $T=0,\Delta=0.197$
(solid line), $T=0.06,\Delta=0.158$ (dashed line), and $T=0.08,\Delta=0.041$
(dot-dashed line).
\label{fig:temp}}
\end{figure}
Decreasing $t'$ means that the length of the remnant Fermi lines
becomes smaller until they shrink to points along the diagonals for $t'=0$
as in a d-wave superconductor. Thus the ratio $W_D/W$ should decrease with 
$t'$. The solid line in Fig.\ref{fig:temp}, 
calculated with $t'=0.05, T=0, n_\sigma=0.48,
1/\tau=0.03$, illustrates that this indeed is the case. Most of the spectral
weight resides now in a very broad and extended high-frequency structure
above $2\Delta \sim 0.35$, only a small part of $W$ is left for the Drude peak.
With increasing temperature the length of the Fermi lines increases and
the spectral weight is shifted from the optical region to the Drude peak,
as shown by the dashed and dot-dashed lines in this figure.   

In conclusion, we have calculated the optical conductivity $\sigma(\omega)$
of a two-dimensional d-CDW conductor taking into account vertex corrections.
The employed formulation can easily be generalized to an arbitrary
symmetry of the CDW state and to a general band structure. We have addressed
and solved the problem of a proper definition of the current to be used in 
calculating $\sigma(\omega)$. Defining the current via the continuity
equation both the renormalization of the one-particle energies due to the
interaction causing the CDW as well as an additional term due to the momentum
dependence of the CDW order parameter enter. Using this renormalized current 
and bare vertices yields a $\sigma(\omega)$ which is shown to be correct 
only in the limit $\omega \rightarrow 0$. On the other hand, using bare 
vertices and the bare electron dispersion violates in general the continuity
equation for the charge density and gives a $\sigma(\omega)$ which is correct
only at large frequencies. The proposed solution is to use bare current
vertices but to include vertex and the corresponding self-energy corrections.
In this way the continuity equation is explicitly fulfilled and the frequency
dependence of the vertex interpolates smoothly from the strongly renormalized
regime at low to the unrenormalized regime at high frequencies. This behavior
is also visible in the numerically calculated conductivity curves. These
calculations show the spectral features of the Drude-like peak at low
and the transitions across the gap at high frequencies and their dependence on 
temperature and the next-nearest neighbor hopping constant $t'$. Remarkable
is also that the temperature dependence of the restricted optical sum $W$
(i.e., the spectral weight integrated over all frequencies) between $T=0$ and 
$T=T_c$ is enhanced by vertex corrections. This may be interpreted as a
violation of the sum rule by about 10 per cents in that interval. This 
as well as the other above predictions may help to identify experimentally   
unconventional CDW's.

\begin{acknowledgments}
The authors would like to thank I.V.\ Gornyi, A. Greco, C. Honerkamp, and M.N.\ 
Kiselev for valuable discussions. \end{acknowledgments}

\end{document}